\documentclass[letterpaper]{article} % DO NOT CHANGE THIS
\usepackage[]{aaai25}  % DO NOT CHANGE THIS
\usepackage{times}  % DO NOT CHANGE THIS
\usepackage{helvet}  % DO NOT CHANGE THIS
\usepackage{courier}  % DO NOT CHANGE THIS
\usepackage[hyphens]{url}  % DO NOT CHANGE THIS
\usepackage{graphicx} % DO NOT CHANGE THIS
\usepackage{natbib}  % DO NOT CHANGE THIS AND DO NOT ADD ANY OPTIONS TO IT
\usepackage{caption} % DO NOT CHANGE THIS AND DO NOT ADD ANY OPTIONS TO IT
\usepackage{algorithm}
\usepackage{algorithmic}

\usepackage{newfloat}
\usepackage{listings}
\DeclareCaptionStyle{ruled}{labelfont=normalfont,labelsep=colon,strut=off} % DO NOT CHANGE THIS
\floatstyle{ruled}
\newfloat{listing}{tb}{lst}{}
\floatname{listing}{Listing}
\title{Improved Large Language Model Jailbreak Detection via Pretrained Embeddings}
\author{
    Erick Galinkin, Martin Sablotny
}
\affiliations{
    \textsuperscript{\rm 1}NVIDIA Corporation\\
    2788 San Tomas Expressway \\
    Santa Clara, CA 95051 USA \\
    egalinkin@nvidia.com
}

\usepackage{bibentry}
\begin{document}

\maketitle

\begin{abstract}
The adoption of large language models (LLMs) in many applications, from customer service chat bots and software development assistants to more capable agentic systems necessitates research into how to secure these systems.
Attacks like prompt injection and jailbreaking attempt to elicit responses and actions from these models that are not compliant with the safety, privacy, or content policies of organizations using the model in their application. 
In order to counter abuse of LLMs for generating potentially harmful replies or taking undesirable actions, LLM owners must apply safeguards during training and integrate additional tools to block the LLM from generating text that abuses the model.
Jailbreaking prompts play a vital role in convincing an LLM to generate potentially harmful content, making it important to identify jailbreaking attempts to block any further steps. 
In this work, we propose a novel approach to detect jailbreak prompts based on pairing text embeddings well-suited for retrieval with traditional machine learning classification algorithms. 
Our approach outperforms all publicly available methods from open source LLM security applications.
\end{abstract}


\section{Introduction}
The growing adoption of large language models, particularly in enterprise applications, has spurred significant research into the security implications of these models and the systems that leverage them.
One area that has received a substantial amount of attention is that of so-called ``jailbreaks'' -- attacks that seek to subvert the safety training of large language models.
Though much of the academic literature is concerned with the generation of offensive or undesirable content~\cite{wei2024jailbroken,mehrotra2023tree,inie2023summon}, the integration of large language models into agentic systems can lead to arbitrary tool usage and the exploitation of vulnerabilities in these agents up to and including remote code execution.
As with many problems in computer science, there are fundamental limitations to safety alignment training~\cite{wolf2023fundamental} such that any sort of attempts to formally constrain latent representations and outputs are intractable -- there always exists some prompt that can elicit the desired behavior.
Defending these systems has evolved along a similar path to other defenses: starting with signature-based detections like string matching and regular expressions, and moving on to heuristics and machine learning-based detections. 
In this work, we couple embedding models with more traditional machine learning techniques to develop jailbreak detection models that outperform all other known, open-source options.

Embedding models serve as a powerful way to map text into a latent space that provides some indication of meaning~\cite{mikolov2013distributed}.
While the overall distribution of words and n-grams between jailbreak prompts and non-jailbreak prompts are quite similar, their embeddings tell a different story.
A simple, na\"ive approach to this problem is the use of vector databases -- embedding a set of known-good and known-jailbreak prompts and using such a database to determine if a user input is likely to be a jailbreak based on its embedding distance from other prompts.
We experiment with this and other approaches, finding that a combination of embeddings and traditional machine learning approaches outperform both pure emdedding distance and pure machine learning approaches.

We begin by covering necessary background, then discuss a variety of approaches that were attempted in conducting this research.
We compare our approach to a variety of models that have been used in both open-source and commercial ``LLM firewalls'', demonstrating our state of the art detection.
Finally, we conclude with limitations of this study and directions for future work.

\section{Background}
\subsection{Adversarial Machine Learning}
Adversarial machine learning dates back to at least 2008, when naive Bayes classifiers used for spam detection were bypassed using machine learning techniques to optimize a message body -- an attack subsequently updated for our modern large language model era by Will Pearce and Nick Landers~\cite{pearce2019proof}.
Modern adversarial machine learning was largely kicked off by Szgedy \textit{et al.}'s 2014 work~\cite{szgedy2014intriguing} on image classification systems, finding optimal perturbations for inducing misclassification in differentiable models, particularly focused on convolutional neural networks.
In large language models, early attacks followed this trend, aiming to be imperceptible~\cite{boucher2022bad} but have since evolved, using similar techniques to those used in computer vision~\cite{zou2023universal}.

In addition to these algorithmically generated attacks, the ``Do Anything Now'' or DAN jailbreaks that use natural language in lieu of an adversarial suffix have spread widely on internet forums like Reddit~\cite{wei2024jailbroken} and many are now using large language models or other natural language processing techniques to automatically generate this style of attack~\cite{mehrotra2023tree}. 
Mitigation and detection of these adversarial attacks has long been a challenge, and though a thorough analysis has not been done for language models, work by Zimmerman \textit{et al.}~\cite{zimmermann2022increasing} found that every proposed defense that has been evaluated was broken by adapting the attack to be aware of the defense.

\subsection{Jailbreak Detection and Mitigation}
At present, a number of approaches to jailbreak detection and mitigation have been attempted, ranging from detection via machine learning classifiers to backtranslation~\cite{wang2024defending} and representation engineering~\cite{zou2024improving}. 
While alignment techniques seek to limit the impact of jailbreaking by mitigating certain behaviors, some research~\cite{wolf2023fundamental} suggests that any intrinsic alignment of the model itself is insufficient, suggesting we should emphasize the role of external controls like guardrails~\cite{rebedea2023nemo}.

Some ``LLM Firewalls'' like the open source Vigil\footnote{https://github.com/deadbits/vigil-llm} have integrated vector databases as an approach to detect and mitigate jailbreak attacks, while others rely on publicly available jailbreak detection models.
This approach has seen growing adoption in industry, as it follows common deployment and usage patterns familiar to cybersecurity practitioners. 
Our work interrogates the effectiveness of such approaches and aims to improve on the current state of the art in the space.

\section{Embedding Models}
Embedding models are used to convert a text (sequence of tokens) to a finite dimensional vector.
Based on the training objective for the embedding model the vector in the latent vector space can have different properties. 
For example, an embedding model that is trained on minimizing distance between similar sequences of tokens provides vectors for similar sequences that are close together under the metric used for training. 
Another example objective is to place question sequences close to their answer sequences under the used metric. 

This work considers four different embedding models released under a variety of licenses -- two of which are permissive, and two of which are Creative Commons non-commercial.
Details on these embedders are given below.

\subsection{\texttt{dpr-ctx\_encoder-single-nq-base} (Meta)}
The Meta embedder is the ``DPR CTX Encoder Single NQ base'' model, a BERT-based model introduced by Facebook (now Meta) in 2020~\cite{karpukhin2020densepassageretrievalopendomain}.
The model was developed for performing retrieval tasks from dense passages.
It has an embedding dimension of 768 and a maximum context length of 512 tokens. 
This model is licensed under Creative Commons non-commercial 4.0.

\subsection{\texttt{NV-Embed-v1} (NVEmbed)}
NV-Embed~\cite{lee2024nvembedimprovedtechniquestraining} is an embedding model based on the Mistral-7B-v0.1~\cite{jiang2023mistral7b} decoder only LLM. 
The embedding model was trained on retrieval and non-retrieval tasks.
It provides an embedding dimension of 4096 and a maximum context length of 4096 tokens. 
This model is licensed under Creative Commons non-commercial 4.0. 

\subsection{\texttt{snowflake-arctic-embed-m-long} (Snowflake)}
The Snowflake model is derived from work by Merrick~\cite{merrick2024embeddingclusteringdataimprove} on unsupervised embeddings.
The embedder has an embedding dimension of 768 and a maximum context length of 8192 tokens.
The model we used, \texttt{snowflake/snowflake-arctic-embed-m-long}, is released under the permissive Apache 2 license.

\subsection{\texttt{nv-embedqa-e5-v5} (NVE5)}
NVE5 is an encoder-only transformer based on the E5-Large-Unsupervised~\cite{wang2022text} model, trained on public datasets.
It has an embedding dimension of 1024 and a maximum context length of 512 tokens. 
This model has been released under the permissive NVIDIA Open Model License. 

\section{Training and Validation Datasets}
\begin{table}[ht]
\begin{tabular}{|l|l|l|l|}
\hline
\textbf{Dataset}         & \textbf{Size}    & \textbf{Jailbreak}    & \textbf{Non-Jailbreak} \\
\hline
\textbf{DAN}             & 16545            & 1405                   & 15140                 \\
\textbf{garak}           & 126              & 126                    & 0                     \\
\textbf{jackhhao}        & 1306             & 666                    & 640                   \\
\hline
\end{tabular}
\caption{Datasets used in aggregated training dataset.} \label{tab:dataset}
\end{table}

The DAN or ``DoAnythingNow'' dataset is sourced from Shen \textit{et al}.~\cite{shen2024characterizing} and contains a number of ``in-the-wild'' jailbreaks -- jailbreak prompts that have demonstrated their efficacy and been shared publicly.
The garak dataset was generated using the \texttt{AutoDAN}~\cite{liu2024autodan} and \texttt{TAP}~\cite{mehrotra2023tree} implementations in the open source \texttt{garak} framework~\cite{derczynski2024garak}.
The jackhhao dataset was sourced from HuggingFace\footnote{https://huggingface.co/jackhhao/jailbreak-classifier}.
The datasets were combined and deduplicated, yielding 17085 examples: 1580 known jailbreaks and 15505 non-jailbreaks broken down in Table~\ref{tab:dataset}.
This aggregate dataset was split into a training (80\%) and validation (20\%) set, stratified by the labels.

\section{Detector Approaches}
We explored four different detector architectures: vector databases, feed forward neural networks, random forests, and XGBoost.
For all model-based approaches -- that is, approaches other than vector databases -- the best hyperparameters for each model were found using grid search. 
All architectures were trained to classify an input to be either a jailbreak or not. 
The following sections introduce the trained architectures in more detail.

\subsection{Vector Databases}
Given the use of embeddings as our source material, vector databases are a natural fit, assuming there is some semantic structure to jailbreak attempts that can be captured by embedding vectors.
Vector databases have proven to be excellent resources for passage retrieval and so-called ``retrieval-augmented generation'' wherein context similar to a provided user input can be retrieved to improve answers to user questions.
We assessed the accuracy, F1 score, and AUPRC for top-1, -3, -5, and -10 matches across both the L2 distance metric and cosine similarity metric using the mean of the results, where jailbreaks are given a positive score and non-jailbreaks are given a negative score for our embeddings. 
The L2 distance metric results proved very poor compared to cosine similarity, and those results have been omitted for brevity.
The best performing vector database configurations were top-1 and top-10 with cosine similarity metrics and were used in subsequent experiments.

\subsection{Neural Networks}
Our neural network classifier was a feed forward neural network with dense layers trained for a maximum of 30 epochs with early stopping.
The input layer was adjusted to fit the embedding dimension and the output layer was set to two dimensions. 
The number of hidden layers and units was freely configurable and configurations from 2 to 8 layers with 2 to 128 neurons per layer were attempted via grid search.
The output of all networks was a 2-dimensional vector used greedily (argmax) to determine whether an input was a jailbreak.
Based on grid search, a 6 layer network with 32 hidden dimensions per layer was the best performing model and was used in subsequent experiments.

\subsection{Random Forests}
Random forests are categorized as an ensemble learning algorithm. 
They combine multiple decision trees to achieve a higher predictive performance than a single decision tree. 
Furthermore, during training they select a subset of features randomly and also apply bagging (bootstrap aggregating), using only a subset of training examples for the decision tree.
The resulting random forest can then be used to classify inputs.
During hyper-parameter tuning we used grid search to vary maximum depths of the trees between 2 and 1024 as well as no depth restriction. 
In addition, we varied the number of estimators between 32 and 1024.
The best performing random forests ultimately used a maximum depth of 20 with 100 estimators and were used in subsequent experiments.

\subsection{XGBoost}
XGBoost~\cite{chen2016xgboost} is a scalable end-to-end decision tree algorithm using a second order Taylor approximation in the loss function to approximate the the Newton-Raphson method for optimization.
The resulting tree can be used to classify inputs according to the learned decision tree.
During hyper-parameter tuning, the tree depth and number of estimators were varied between 2 and 2048 via grid search.
The optimal hyperparameters for XGBoost were found to be a depth of 2 with 2048 estimators and this model was used in subsequent experiments.

\section{Results} 
Overall, we found that for all embeddings other than NVEmbed embeddings, where the neural network was best-performing, random forests provided the best performing detectors as determined by 5-fold cross validation.
We further found that the best performing embedding model for vector databases was \texttt{NV-Embed-v1}.
For brevity, we provide details for the best performing configurations of embeddings and detectors on our evaluation set in Tables~\ref{tab:xval} and \ref{tab:results} followed by a detailed analysis of the best performing detectors for each embedding against open models on a number of publicly available datasets in the following section.
Given the imbalance in all known jailbreak datasets, we opt to report F1 score in lieu of accuracy metrics throughout this work.

\subsection{Detector Results}
\begin{table*}[ht]
\centering
\begin{tabular}{|l|l|l|l|l|l|l|}
\hline
\textbf{Detector}                                                   & \textbf{Fold 1} & \textbf{Fold 2} & \textbf{Fold 3} & \textbf{Fold 4} & \textbf{Fold 5} & \textbf{Average} \\
\hline
\begin{tabular}[c]{@{}l@{}}Random Forest\\ (Meta)\end{tabular}      & 0.7018          & 0.9371          & 0.8485          & 0.8790          & 0.9366          & 0.8606           \\
\hline
\begin{tabular}[c]{@{}l@{}}Neural Network\\ (NVEmbed)\end{tabular}  & 0.7902          & 0.8609          & 0.8605          & 0.7390          & 0.8647          & 0.8231           \\
\hline
\begin{tabular}[c]{@{}l@{}}Random Forest\\ (Snowflake)\end{tabular} & 0.9912          & 0.9910          & 0.9938          & 0.9947          & 0.8966          & \textbf{0.9735}  \\
\hline
\begin{tabular}[c]{@{}l@{}}Random Forest\\ (NVE5)\end{tabular}      & 0.9848          & 0.9835          & 0.9964          & 0.9936          & 0.8950          & 0.9707           \\
\hline
\begin{tabular}[c]{@{}l@{}}VectorDB k=1\\ (NVEmbed)\end{tabular}    & 0.7582          & 0.8298          & 0.8644          & 0.7357          & 0.7788          & 0.7934           \\
\hline
\begin{tabular}[c]{@{}l@{}}VectorDB k=10\\ (NVEmbed)\end{tabular}   & 0.7361          & 0.2808          & 0.8202          & 0.6753          & 0.7779          & 0.7662           \\  
\hline
\end{tabular}
\caption{F1 Scores for detectors under 5-fold cross-validation. Higher scores are better.} \label{tab:xval}
\end{table*}

\begin{table}[h!]
\begin{tabular}{|l|l|l|l|l|}
\hline
\textbf{Detector}                                                   & \textbf{F1 Score}  & \textbf{FPR}     & \textbf{FNR}     \\
\hline
\begin{tabular}[c]{@{}l@{}}Random Forest \\ (Meta)\end{tabular}     & 0.8289             & 0.0089           & 0.2313           \\
\hline
\begin{tabular}[c]{@{}l@{}}Neural Network\\ (NVEmbed)\end{tabular}  & 0.8237             & 0.0236           & 0.1313           \\
\hline
\begin{tabular}[c]{@{}l@{}}Random Forest\\ (Snowflake)\end{tabular} & \textbf{0.8333}    & 0.0092           & 0.2188           \\
\hline
\begin{tabular}[c]{@{}l@{}}Random Forest\\ (NVE5)\end{tabular}      & 0.8243             & \textbf{0.0086}  & 0.2375           \\
\hline
\begin{tabular}[c]{@{}l@{}}VectorDB k=1\\ (NVEmbed)\end{tabular}    & 0.7934             & 0.0313           & \textbf{0.1196}  \\
\hline
\begin{tabular}[c]{@{}l@{}}VectorDB k=10\\ (NVEmbed)\end{tabular}   & 0.7540             & 0.0413           & 0.1257           \\
\hline
\end{tabular}
\caption{Detector F1 scores, false positive rate (FPR), and false negative rate (FNR). Higher scores for F1 are better, lower scores for FPR and FNR are better.} \label{tab:results}
\end{table}

The six detectors assessed were chosen based on the highest average F1 score achieved in 5-fold cross validation, with results for our best performing detectors shown in Table~\ref{tab:xval}.
For each of the four embeddings, the highest performing model was chosen.
For the vector databases, the two best performing combinations of embeddings and $k$ values were chosen. 

Table~\ref{tab:results} shows that our best performing detector overall in terms of F1 score was the Random Forest with Snowflake embeddings (0.8333).
Notably, all detectors were highly competitive with one another, with only 0.0753 difference in F1 score between our highest and lowest performing detector. 
Curiously, the NVEmbed embeddings were the best performing in terms of F1 score for our vector databases, and the same embeddings resulted in the best performing model being a neural network, while all other embeddings had a random forest as their best performing model architecture.

\subsection{Comparison to Public Models}
A number of approaches have been attempted to detect and mitigate LLM jailbreaks. 
Analogously to malicious network traffic detection, some of these are based on static rules akin to \texttt{iptables} while other approaches leverage machine learning.
Many so-called ``LLM firewalls'' incorporate one or more of these models.
We opt to compare our two most permissively licensed approaches to three of these public, permissively licensed models, two of which are sourced from popular, open source LLM firewall applications: \texttt{gelectra-base-injection} (gelectra)\footnote{\url{https://huggingface.co/JasperLS/gelectra-base-injection}}, \texttt{deberta-v3-base-injection} (deberta)\footnote{\url{https://huggingface.co/JasperLS/deberta-v3-base-injection}}, and PromptGuard~\footnote{\url{https://www.llama.com/docs/model-cards-and-prompt-formats/prompt-guard/}}. 
Given the popularity of using powerful language models like gpt-3.5-turbo to perform these sorts of checks, we also consider this approach, using the \texttt{gpt-3.5-turbo-0125} model and a prompt from the open source Rebuff\footnote{\url{https://github.com/protectai/rebuff}} project. 
We run our comparison over three evaluation sets: our own test set, ToxicChat, and JailbreakHub.

\begin{table}[h!]
\begin{tabular}{|l|l|l|l|}
\hline
\textbf{Detector}                                                   & \textbf{F1 Score}  & \textbf{FPR}     & \textbf{FNR}     \\
\hline
\begin{tabular}[c]{@{}l@{}}Random Forest\\ (Snowflake)\end{tabular} & \textbf{0.8333}             & 0.0092           & 0.2188           \\
\hline
\begin{tabular}[c]{@{}l@{}}Random Forest\\ (NVE5)\end{tabular}      & 0.8243             & \textbf{0.0086}           & 0.2375           \\
\hline
\begin{tabular}[c]{@{}l@{}}gelectra\end{tabular}                    & 0.2046             & 0.7199           & 0.0500           \\
\hline
\begin{tabular}[c]{@{}l@{}}deberta\end{tabular}                     & 0.2173             & 0.6745           & \textbf{0.0406}           \\
\hline
\begin{tabular}[c]{@{}l@{}}PromptGuard\end{tabular}                 & 0.4518             & 0.1972           & 0.1219           \\
\hline
\begin{tabular}[c]{@{}l@{}}gpt-3.5-turbo\end{tabular}               & 0.2438             & 0.0313           & 0.1688           \\
\hline
\end{tabular}
\caption{Comparisons over our own test set. Higher scores for F1 are better, lower scores for FPR and FNR are better. The best performer in each category has been bolded.} \label{tab:eval}
\end{table}

Our models outperform all others on our aggregated test set, shown in Table~\ref{tab:eval}. 
While this is expected given that the test set is drawn from the same data distribution as our training data, our best performing model outperforms the next closest competitor, PromptGuard, by a significant margin in terms of F1 score.
In terms of false positive rate -- an attribute that can have severe impacts on the usability of a system -- all BERT-based detectors have an unacceptably high FPR, even when using the very high threshold for classification: 0.9 for deberta and 0.98 for gelectra.

\begin{table}[h!]
\begin{tabular}{|l|l|l|l|l|}
\hline
\textbf{Detector}                                                   & \textbf{F1 Score}  & \textbf{FPR}     & \textbf{FNR}     \\
\hline
\begin{tabular}[c]{@{}l@{}}Random Forest\\ (Snowflake)\end{tabular} & 0.2833             & \textbf{0.0024}           & 0.8132           \\
\hline
\begin{tabular}[c]{@{}l@{}}Random Forest\\ (NVE5)\end{tabular}      & 0.3325             & 0.0461           & 0.2967           \\
\hline
\begin{tabular}[c]{@{}l@{}}gelectra\end{tabular}                    & 0.0662             & 0.5026           & \textbf{0.0220}           \\
\hline
\begin{tabular}[c]{@{}l@{}}deberta\end{tabular}                     & 0.0845             & 0.3860           & \textbf{0.0220}           \\
\hline
\begin{tabular}[c]{@{}l@{}}PromptGuard\end{tabular}                 & \textbf{0.5287}             & 0.0204           & 0.1868           \\
\hline
\begin{tabular}[c]{@{}l@{}}gpt-3.5-turbo\end{tabular}               & 0.0888             & 0.3395           & 0.0879           \\
\hline
\end{tabular}
\caption{Comparisons over ToxicChat dataset. Higher scores for F1 are better, lower scores for FPR and FNR are better. The best performer in each category has been bolded.} \label{tab:toxicchat}
\end{table}

ToxicChat~\cite{lin2023toxicchat} is not a dataset of jailbreaks but contains many of them, along with many undesirable non-jailbreak prompts.
Although in our review of the dataset the actual number of jailbreaks is slightly lower than the 226 in the 10164 entries of the dataset, the numbers in Table~\ref{tab:toxicchat} reflect the labels included in the HuggingFace version of the dataset\footnote{\url{https://huggingface.co/datasets/lmsys/toxic-chat}} and not our analysis.
PromptGuard performs best on this dataset due to the high false negative rate of our classifiers.
We note that gelectra, deberta, and gpt-3.5-turbo all have a false positive rate greater than 1/3 -- again, unacceptably high for production usage.

\begin{table}[h!]
\begin{tabular}{|l|l|l|l|l|}
\hline
\textbf{Detector}                                                   & \textbf{F1 Score}  & \textbf{FPR}     & \textbf{FNR}     \\
\hline
\begin{tabular}[c]{@{}l@{}}Random Forest\\ (Snowflake)\end{tabular} & \textbf{0.9601}             & \textbf{0.0042}           & 0.0435           \\
\hline
\begin{tabular}[c]{@{}l@{}}Random Forest\\ (NVE5)\end{tabular}      & 0.6188            & 0.1426           & \textbf{0.0030}           \\
\hline
\begin{tabular}[c]{@{}l@{}}gelectra\end{tabular}                    & 0.1926            & 0.9729           & \textbf{0.0030}           \\
\hline
\begin{tabular}[c]{@{}l@{}}deberta\end{tabular}                     & 0.1947            & 0.9596           & \textbf{0.0030}           \\
\hline
\begin{tabular}[c]{@{}l@{}}PromptGuard\end{tabular}                 & 0.3029            & 0.5066           & 0.0450           \\
\hline
\begin{tabular}[c]{@{}l@{}}gpt-3.5-turbo\end{tabular}               & 0.2124            & 0.7207           & 0.1456           \\
\hline
\end{tabular}
\caption{Comparisons over JailbreakHub dataset. Higher scores for F1 are better, lower scores for FPR and FNR are better. The best performer in each category has been bolded.} \label{tab:jailbreakhub}
\end{table}

JailbreakHub~\cite{shen2024characterizing} is a dataset comprised of in the wild jailbreaks and benign prompts sourced from Reddit, Discord, and other places where such prompts are shared.
As such, it is the dataset that seems to best represent the actual jailbreak attempts likely to be used against a target system.
F1 score, FPR, and FNR for all models are shown in Table~\ref{tab:jailbreakhub}.
On this dataset, our random forest detector trained on Snowflake embeddings achieves an F1 score (0.9601) more than three times the highest publicly available model, PromptGuard (0.3029).
The significant improvement of our model compared to the best performing public model underscores the value of our approach -- achieving a high rate of detection while having a false positive rate below 1\%.
Crucially, we point out that all other models including the best-performing public model, PromptGuard, and the most powerful model, gpt-3.5-turbo, have a false positive rate exceeding 50\%.

\section{Discussion}
Our results show a strong performance gain of our novel approach for detecting jailbreak attempts in prompts over existing techniques. 
In the most realistic data set (JailbreakHub), we were able to outperform the second best by more than three times in terms of F1 score with a similar but lower false negative rate. 
Furthermore, the false positive rate was reduced to a fraction compared to PromptGuard (\ref{tab:jailbreakhub}). 
The results suggest that training a general question-answer embedding model first followed by a classifier on the embedding outputs performs better than the end-to-end approach used in the BERT-based approaches (gelectra, deberta, and PromptGuard). 
One reason for the performance difference could be the training itself. 
The BERT-based approaches are all fine-tuned after the initial training to classify jailbreaks. 
One interesting direction for future research is to test the performance of the BERT-based models by adding classification layers to them and only train those for jailbreak detection, freezing the embedding layers. 
This could be contrasted with experiments in which the embedding models and classifier are trained end-to-end, meaning the weights of the embedding layers are also fine-tuned according the the jailbreak detection task.

The results also suggest that a general model such as gpt-3.5-turbo performs worse despite a larger number of total parameters, highlighting the need for specialized techniques to detect jailbreak attempts.
While adoption of techniques like LLM-as-a-Judge have been fruitful in many tasks, it appears that general purpose large language models are still not very effective at identifying jailbreak attempts.
Furthermore, comparing the performance to a general purpose LLM, it is also important to note the difference in both monetary cost for deployment and latency. 
While these numbers were not analyzed in this report, the lower number of parameters in our models generally correlates with fewer resources on all fronts. 

The NVEmbed model was our best performing embedder in our vector database detectors and the only model that performed better in combination with a neural network classifier, in contrast to the other embeddings that performed all better using a random forest classifier. 
One possible reason might be the embedding dimensions in the NVEmbed model. 
It embeds the input into a 4096-dimensional space, which is four times the size of the next smaller embedding space. 
The second reason could be the general training and setup of the models. 
The NVEmbed model is the only model based on a general-purpose LLM in contrast to the other embedding models.  
Those are either trained from end to end as an embedding model (Meta, Snowflake) or, in the NVE5 case, fine-tuned versions of a model trained from end to end as an embedding model.
Further research is needed to identify the exact reason for these differences.

\section{Conclusion}
This work introduces a novel approach to identify jailbreaks by pairing high quality embedding models intended for use in retrieval systems with traditional machine learning classification algorithms.
Overall, we find that our technique is a significant improvement for hardening LLM deployments against jailbreak attacks and can be easily integrated into a variety of guardrail and LLM firewall type systems.
We demonstrate significant improvements over existing, publicly available methods, particularly on the realistic JailbreakHub dataset. 
The combination of the Snowflake embeddings and a random forest classifier yielded the best results overall across our own evaluation set and a public evaluation set, outperformed in terms of F1 score only by one model on one public evaluation set.

There are limitations to our work.
Although results on unseen prompts suggest a degree of generalizability, only deployments in production environments can provide sufficient data to evaluate long term performance, even ignoring model drift. 
We note, however, that since our approach does not modify the embedding models and uses a relatively lightweight classifier in the form of a random forest, the cost of retraining to account for model drift is significantly lower than in transformer-based approaches. 
In future work, we aim to explore the potential performance impacts of fine tuning the embedding models during classifier training compared to our method.

\bibliography{aaai25}

\end{document}